\documentclass[preprint, aps, prb, longbibliography]{revtex4-1}

 \pdfoutput=1

\usepackage{graphicx}
\usepackage{color}
\usepackage{epsfig}
\usepackage{subfigure}
\usepackage{natbib}
\usepackage{array}
\usepackage{amsmath}
\usepackage{enumitem} 

\usepackage{lineno}

\makeatletter

\begin{document}
\newcolumntype{M}{>{$}c<{$}}

%

\title{A new class of intrinsic magnet: two-dimensional yttrium sulphur selenide} 

\author{Pankaj Kumar, Ivan I. Naumov, Priyanka Manchanda and Pratibha Dev}

\affiliation{Department of Physics and Astronomy, Howard University, Washington, D.C. 20059, USA}
\date{\today}

%
%


\begin{abstract}

	Exploring and controlling magnetism in two-dimensional (2D) layered magnetic crystals, as well as their inclusion in heterogeneous assemblies, provide an unprecedented opportunity for fundamental science and technology. To date, however, there are only a few known intrinsic 2D magnets. Here we predict a novel 2D intrinsic magnet, yttrium sulphur selenide (YSSe), using first principles calculations. The magnetism of this transition metal dichalcogenide originates from the partially-filled $3p$- and $4p$-orbitals of the chalcogens, unlike other known intrinsic magnets where magnetism arises from the partially-filled $3d$- and $4f$-orbitals. The unconventional magnetism in YSSe is a result of a unique combination of its structural and electronic properties. We further show that a lack of mirror symmetry results in piezoelectric properties, while the broken space- and time-symmetry ensures valley polarization. YSSe is a rare magnetic-piezoelectric material that can enable novel spintronics, valleytronics and quantum technologies.
\end{abstract}
\maketitle

Until recently, long-ranged magnetic ordering was believed to be impossible in 2D layered materials due to its destruction by increased thermal agitations, in accordance with the Mermin-Wagner Theorem. The discovery of long-ranged intrinsic magnetism in pristine 2D materials---$\mathrm{CrI_{3}}$~\cite{Huang_CrI3_2017} and $\mathrm{Cr_{2}Ge_{2}Te_{6}}$~\cite{Gong2017}---showed that magnetism can survive in 2D due to the preferred alignment direction of the local moments (magnetic anisotropy), which counters thermal excitations.  Magnetism in 2D crystals has inspired a renewed interest in the fundamental physics of long-ranged magnetism in reduced dimensions. This discovery also offers technological opportunities, with potential uses in miniaturized and flexible spintronics devices, as well as proximity-effect devices, such as the superconductor/semiconductor/magnet heterostructure that was proposed by Sau \textit{et al.}~\cite{Sau_DasSarma_PRL_2010} as the solid-state platforms for Majorana bound states. In spite of an active search for additional magnetic 2D crystals, there are very few known intrinsic 2D magnets. Here we present our theoretical discovery of a yttrium-based magnetic semiconductor, yttrium sulphur selenide (YSSe), which  is a Janus transition metal dichalcogenide. The word ``Janus" is used to emphasize that the structure lacks mirror symmetry, with two different chalcogens (S and Se) on the two opposite faces. The magnetism in YSSe is shown to originate from the unpaired electrons in \textit{3p}- and \textit{4p}-derived states of the chalcogens. To the best of our knowledge, this is the first time an intrinsic magnet is reported with moments contributed by \textit{3p}- and \textit{4p}-derived states, although these orbitals are known to be involved in defect-induced magnetism in otherwise non-magnetic $\mathrm{PtSe_{2}}$~\cite{PtSe2_Avsar_2019}.  In addition to the broken time-reversal symmetry, YSSe lacks inversion symmetry as well as mirror symmetry. The unique combination of properties, which ensure that YSSe is both a piezoelectric and a magnet, can potentially open doors to new applications. 

\vspace{12pt}
\noindent \textbf{Results}

\vspace{8pt}

\begin{figure*}[ht]
	\begin{center}
	\includegraphics[width=0.85\linewidth]{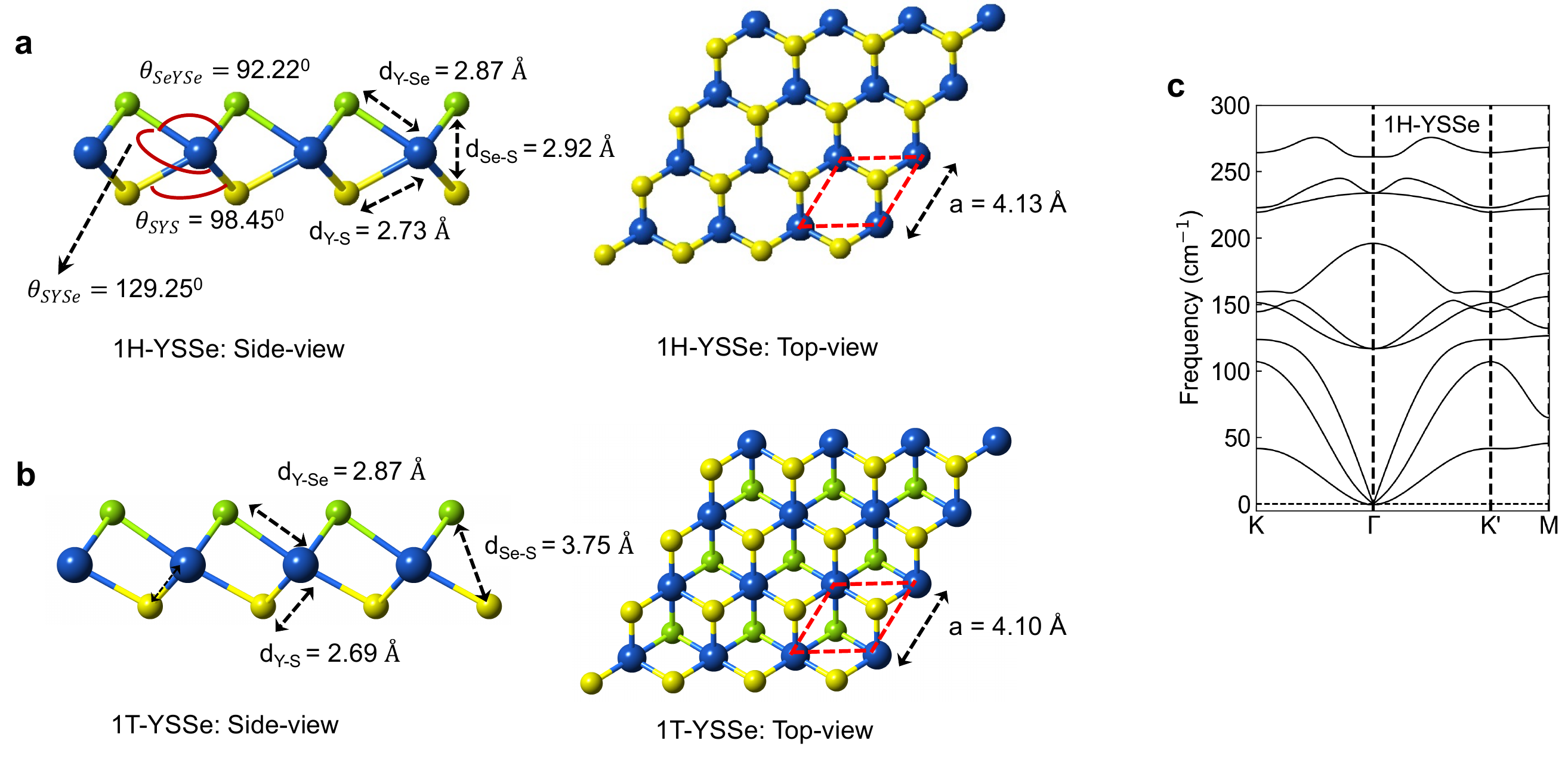}
	 \end{center}
 \vspace{-12pt}
	\caption{\label{fig:image1} Structural properties of YSSe monolayers: \textbf{a} Side- and top-views of 1H-YSSe. \textbf{b} Side- and top-views of 1T-YSSe. \textbf{c} Phonon dispersion for 1H-YSSe showing no negative frequencies, demonstrating dynamical stability of the predicted structure.}
\end{figure*}


\noindent \textbf{Structural, electronic and spin properties of YSSe monolayers.}  Our calculations are based on density-functional theory (DFT) within the generalized gradient approximation (GGA)~\cite{GGA}, using the Vienna \textit{Ab-Initio} Simulation package (see Methods section). YSSe exists in two low-energy phases: (i) a semiconducting 1H-phase, which we find to be magnetic, and (ii) a non-magnetic metallic 1T-phase. For the semiconducting 1H-phase, seen in Figure~\ref{fig:image1}(a), the Y-Se bond is longer than the Y-S bond and the bond-angles -- $\theta_{SYS}$ and $\theta_{SeYSe}$ -- are also different, resulting in an asymmetrical structure with $C_{3v}$-symmetry (broken trigonal prismatic symmetry). As shown in Figure~\ref{fig:image1}(b), 1T-YSSe also has $C_{3v}$-symmetry (broken octahedral-symmetry) due to the lack of a mirror plane in a Janus TMD and anisotropic bonding. The non-magnetic, metallic 1T-phase is lower in energy by $9.07\,meV$ as compared to the 1H-phase, and both structures are likely to occur at room temperature.   As we are interested in the magnetic structure, we will concentrate on the 1H-phase, making comparisons with the 1T-phase when needed.
 
 

Since we are predicting a new material, we studied its inter-dependent structural, electronic and spin properties. The calculated phonon dispersion for 1H-YSSe, shown in Figure~\ref{fig:image1}(c), has no negative phonon branches, demonstrating dynamical stability of the predicted structure. Furthermore, our \textit{ab initio} dynamics simulations, performed at $300\,K$ for $10\,ps$, show that the structure is thermodynamically stable. The cohesive energy of YSSe ($4.722\,eV/atom$) in the 1H-phase is large, where the cohesive energy, $E_{C}$ is defined as the energy required to separate a crystal into constituent isolated atoms. It is calculated using the formula: $E_{C}=-(E_{YSSe} - E_Y - E_S -E_{Se})/3$, with $E_{X}$ being the total energy of species $X$.

\begin{figure*}[ht]
	\centering
	\hspace{-0.5cm}
	\includegraphics[width=0.85\linewidth]{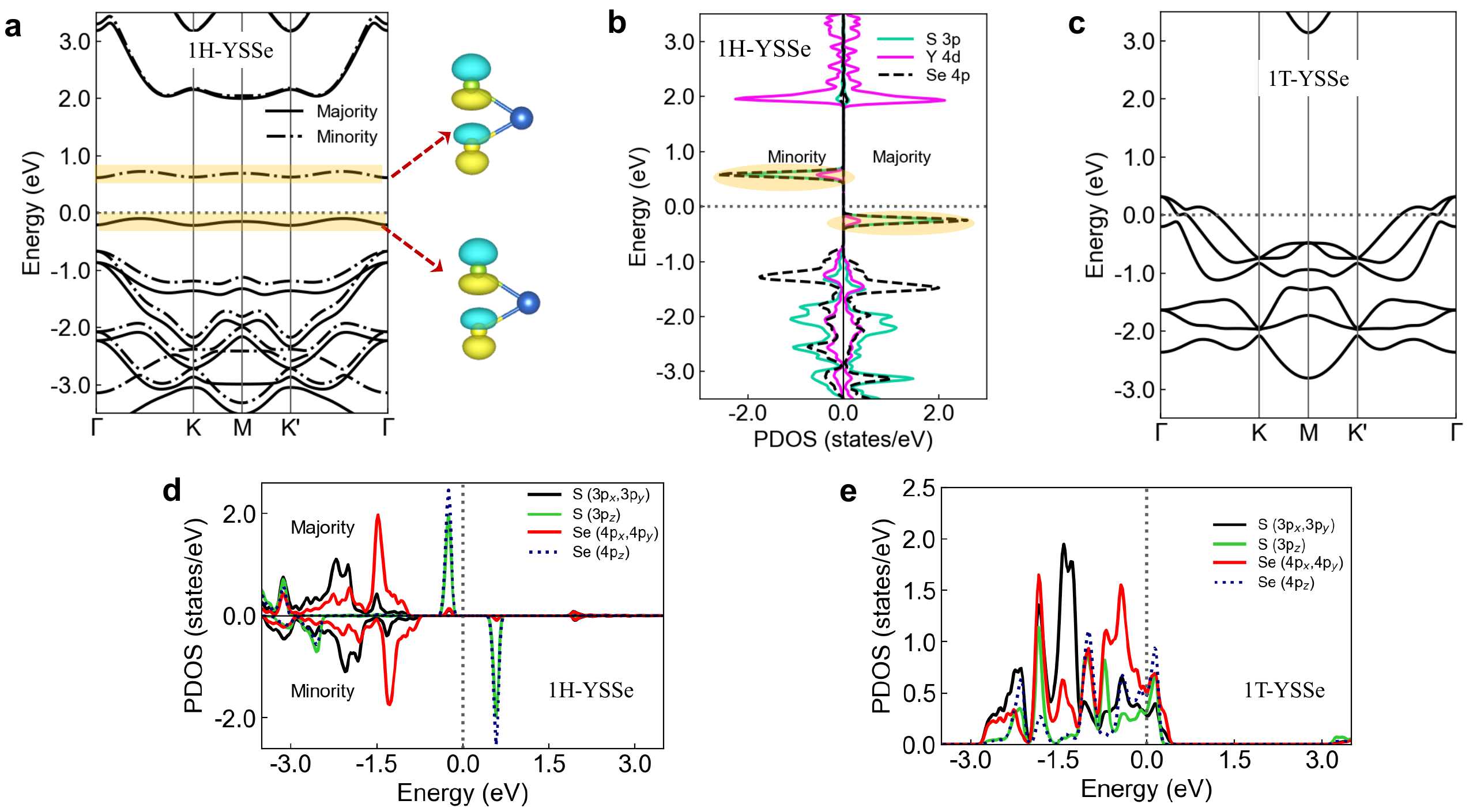}
	\caption{Electronic properties of YSSe monolayers: \textbf{a}  Band structure for 1H-YSSe showing nearly-dispersionless, spin polarized bands (highlighted) for the filled majority-spin and the empty minority-spin channels. Also plotted are charge density isosurfaces (yellow and blue for positive and negative isovalues), showing that these bands originate in the \textit{3p}- and \textit{4p}-derived states of S and Se, with antibonding character. \textbf{b} Projected density of states plotted for the $p$-states of S and Se, along with $d$-states of Y in 1H-YSSe, showing sharp peaks around the Fermi level (reference energy). \textbf{c} Band structure for the metallic 1T-YSSe. \textbf{d} and \textbf{e}  Projected density of states for 1H-YSSe and 1T-YSSe with components of p-orbitals of S and Se,  showing the switching of the order of the $p_{z}$-derived singlet state, and the doublet-states composed of the $p_x$ and $p_y$ of S and Se between the two phases.}
	\label{fig:image2}
\end{figure*}

 The electronic structure and spin properties of 1H-YSSe can be understood within Crystal Field Theory, taking the electronic configuration of yttrium ([Kr] 4\textit{d}$^1$ 5\textit{s}$^2$), which is a group-III element, into account. In contrast to the $+4$ formal oxidation state of the transition elements in group-VI based TMDs, such as MoS$_2$, or the synthesized, non-magnetic Janus structure, MoSSe~\cite{MoSSe_Zhang_2017}, yttrium adopts a $+3$ formal oxidation state in YSSe. This leaves the S- and Se-atoms in YSSe under-coordinated, with a total of one unpaired electron, which is almost equally split between the two chalcogens due to their near-equal electronegativities, $\chi$ ($\chi_{S}=2.58$ and $\chi_{Se}=2.55$).  Furthermore, due to the $C_{3v}$-symmetry of 1H-YSSe, the triply-degenerate $p$-orbitals of the chalcogens split into a doublet ($p_{x}$, $p_{y}$) and a singlet ($p_{z}$). 
In the case of 1H-YSSe, due to the smaller distance between the S and Se atoms [$d_{Se-S}=2.92$\,\AA{}, see Figure~\ref{fig:image1}(a)], the $3p_{z}$ and $4p_{z}$ states of S and Se, respectively, which overlap in space and energy, hybridize and form bonding $A_{1}$ and antibonding $A_{1}^*$-states. 
The antibonding $A_{1}^*$-state in 1H-YSSe is closest to the valence band edge. This is in direct contrast with the 1T-YSSe phase structure, where $d_{Se-S}=3.75$\,\AA{}, resulting in smaller splitting between the bonding $A_{1}$ and antibonding $A_{1}^*$-states, with the latter being lower in energy than the doublet states, $E$. The $E$-states are partially filled and are composed of hybridized doublets, derived from yttrium's $d$-states and the p-derived ($p_{x}$, $p_{y}$) states of S and Se.  Figure~\ref{fig:image2}(a) shows the resulting spin-resolved band structure of a 1H-YSSe monolayer along the high-symmetry points of the Brillouin zone ($\Gamma$-$K$-$M$-$K^{\prime}$-$\Gamma$).  The band structure shows spontaneous spin splitting with more electrons occupying one of the spin channels (majority spin) versus the other (minority-spin channel) due to the exchange interaction. The large exchange splitting between majority and minority spin $A_{1}^*$-states is due to quantum-confinement effects in the 2D-crystal, along with the structural attributes of 1H-YSSe. The large covalent radius of a yttrium atom ($1.90$\,\AA{}) results in a large separation between neighboring atoms on the two faces of the TMD ($d_{S-S}=d_{Se-Se}=4.13$\,\AA{}). The spatial localization, in turn, results in a large exchange interaction. This spatial localization is indicated by the nearly dispersionless $A_{1}^*$ majority- and minority-spin bands [highlighted in Figure~\ref{fig:image2}(a)], with bandwidths of $\sim0.12\,eV$ and $\sim0.11\,eV$, respectively. Figure~\ref{fig:image2}(a) also shows the charge density plots of the $A_{1}^*$-state for the majority and minority spin channels at the $\Gamma$-point, with yellow (blue) corresponding to positive (negative) isosurfaces. The isosurface plots show that the $p_{z}$-derived states of the chalcogens are very localized, retaining the atomic-like character of p-orbitals even in the solid state. This localization and spin-splitting is also evident in the projected density of states plotted in Figure~\ref{fig:image2}(b). This also explains why the 1H-phase of YSSe is semiconducting, while its 1T phase is metallic. 1H-YSSe is a magnetic semiconductor. It has a majority-spin $A_{1}^*$-state that is completely filled and a minority-spin $A_{1}^*$-state that is completely empty. On the other hand, 1T-YSSe, with the partially filled $E$-states, is a (non-magnetic) metal as can be seen in its band structure plotted in Figure~\ref{fig:image2}(c). Figures~\ref{fig:image2}(d)-(e) plot the component-resolved projected density of states, showing the switching of the order of the singlet state, composed of $p_{z}$ of S and Se, and the doublet, composed of $p_x$ and $p_y$ of S and Se, between the two phases. The electronic structure of 1H-YSSe leads to its calculated properties, including: (i) the formation of a local magnetic moment of $1\mu_{B}$ per formula unit, most of which comes from S ($\sim0.45\,\mu_{B}$) and Se ($\sim0.56\,\mu_{B}$), with a small counter-polarized moment coming from Y ($\sim-0.02\,\mu_{B}$), and (ii) an indirect band gap of 2.094\,eV in the majority spin-channel, with the valence band maximum lying along the $\Gamma-K$ direction and the conduction band maximum at the $M$-point, and a direct bandgap of 1.28\,eV in the minority spin-channel at the $\Gamma$-point.



\vspace{12pt}

\noindent \textbf{Long-ranged magnetism in 1H-YSSe.} So far, we have shown that the 1H-YSSe has a local magnetic moment. However, in 2D crystals, collective magnetism, i.e. long-ranged magnetism, survives only if the crystals display a preferred alignment direction of the local moments (magnetic anisotropy), which counters the destructive effects of thermal excitations. Magnetic anisotropy originates from spin-orbit coupling (SOC) and can be quantified by calculating the magnetic anisotropy energy ($MAE$), which is defined as the difference in total energies of the structures with magnetizations that are parallel and perpendicular to the atomic plane: $MAE=E_{\parallel}-E_{\perp}$. Here, $E_{\parallel}$ and $E_{\perp}$ are obtained from noncollinear DFT calculations with fully-relativistic pseudopotentials that take SOC into account. We find that 1H-YSSe has an easy magnetization plane (XY) with an MAE value of about $-122\,\mu$\,eV. Hence, YSSe is an XY-magnet with no energy barrier to the rotation of the moment in the atomic plane, thereby, exhibiting continuous O(2) spin symmetry.  In accordance with the Mermin-Wagner theorem, for 2D YSSe with an O(2) spin-symmetry, there should be no long-ranged ordered state at finite temperatures.  Nevertheless, the system is allowed to undergo a Berezinskii-Kosterlitz-Thouless (BKT) transition  from a high-temperature disordered phase to a low-temperature, quasi-ordered phase. The critical temperature, $T_{C}$, of the BKT transition in an XY-magnet can be estimated by combining  the results from DFT calculations to those from Monte Carlo simulations of the XY model~\cite{XY_MonteCarlo_1986, Hennig_XY_2017}, according to which $T_{C}=0.89J/k_{B}$. Here,  $k_{B}$ is the Boltzmann constant and $J$ is the exchange coupling between magnetic moments. $J$ can be obtained from the difference between the total energies of YSSe with moments aligned in ferromagnetic (FM) and antiferromagnetic (AFM) configurations: $\Delta E_{mag}=E_{AFM}-E_{FM}=8\,J\,\mu^2$, where $\mu$ is the magnetic moment per formula unit. We find that  $\Delta E_{mag}$ is $+2.90$\,meV per formula unit in a collinear (i.e. non-relativistic) calculation, while the non-collinear calculations (including SOC) yield a $\Delta E_{mag}$ of $+$3.07\,meV per formula unit. The positive sign of the calculated $\Delta E_{mag}$ means that FM ordering is the preferred ordering of moments. Using the calculated values of $\Delta E_{mag}$ in $T_{C}=0.89\Delta E_{mag}/8\mu^{2}k_{B}$, we estimate the critical temperature to be $T_{C}=3.74$\,K (without SOC) and a $T_{C}=3.96$\,K with SOC effects taken into account.  

\begin{figure*}[ht]
	\centering
	\hspace{-0.5cm}
	\includegraphics[width=0.95\linewidth]{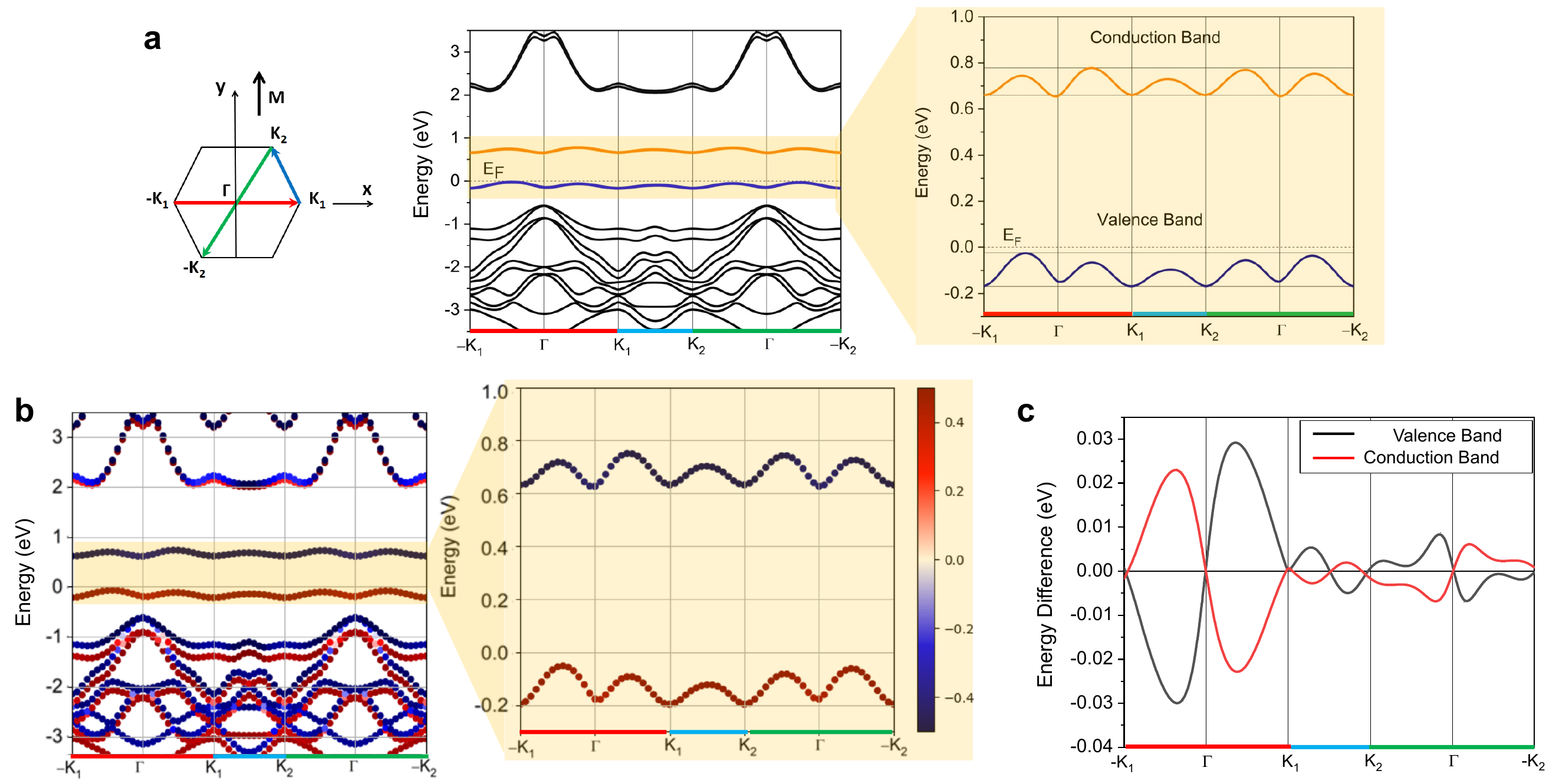}
	\caption{ Effects of SOC: \textbf{a} Band structure along the paths indicated by red, blue and green lines in the Brillouin zone (BZ) shown on the left. The magnetization, $\textbf{\textit{M}}$, is taken to be parallel to the $y$-direction, corresponding to the C$_{1h}$(C$_{1}$) magnetic group. This lowering of symmetry results in a larger irreducible Brillouin zone, with the $ K_{1}$ and $ K_{2}$ being distinct due to different projections of $K_{1}\to \Gamma$ and $K_{2}\to \Gamma$ along $\textbf{\textit{M}}$. The highlighted region of the band structure shows only the topmost valence and lowest conduction bands. \textbf{b} Spin-resolved band structure for 1H-YSSe along different paths in the BZ for the case when the magnetization is directed along the $y$ direction. Only the  $\pm S_y$ spin projections are shown since  $\pm S_x$ and $\pm S_z$  projections are almost negligible. The highlighted region of the spin-resolved band structure shows that the topmost valence band is mostly spin up and lowest conduction band is mostly spin down. The colors are used to encode the expectation values of the projections. \textbf{c} The magnetic anisotropy energy contributions calculated  along the $\boldsymbol{k}$-lines in the BZ. Only the contributions corresponding to the highest valence and lowest conduction bands are presented.}
	\label{fig:image3}
\end{figure*}
In order to understand the small values of $\Delta E_{mag}$ and hence, the exchange coupling observed in our DFT calculations for YSSe, we turn to and expand the scope of the semiempirical Goodenough-Kanamori rules for superexchange interactions.  The superexchange mechanism refers to exchange coupling between the two nearest magnetic atoms with partially filled orbitals, mediated through an intermediary, non-magnetic atom. Conventionally, the magnetic ions are transition metal cations, bridged by an anion (e.g. O$^{2-}$ in MnO). In YSSe, it is the S and Se atoms that have the partially filled orbitals and carry the magnetic moments, and the exchange coupling  between the magnetic anions is mediated by the non-magnetic cation, Y. In spite of this role reversal, the superexchange mechanism can be applied to understand exchange coupling in YSSe. This is because, in principle, there is no physical law against: (i) the formation of local moments due to partially filled localized orbitals in an anion, and (ii) mediation of exchange through an intermediate non-magnetic cation. From the Goodenough-Kanamori rules, we know that superexchange maximally favors AFM (FM) coupling if the anion-cation-anion angle is $180^{\circ}$ ($90^{\circ}$), maximizing the hopping between the magnetic ions. The small values of $\Delta E_{mag}$ with and without SOC are a result of the competition between AFM coupling, which is favored between moments on S and Se [with $\theta_{SYSe}=129.25^{\circ}$ in Figure~\ref{fig:image1}(a)], and FM coupling, which is favored between moments on S (Se)-atoms on the same faces [with $\theta_{SYS}=98.45^{\circ}$ and $\theta_{SeYSe}=92.22^{\circ}$ in Figure~\ref{fig:image1}(a)]. In order to prove that the proposed exchange mechanism is indeed dictating the long-range magnetic behavior, and in turn, the calculated value of $\Delta E_{mag}$, we studied the effect of hydrostatic strain on this quantity.  The applied strain changed the angles between the overlapping orbitals, affecting the $\Delta E_{mag}$-values in keeping with the Goodenough-Kanamori rules [see Supplementary Figure 1].


We further analyzed the effects of SOC by performing band structure calculations along several  $\boldsymbol{k}$-lines in the Brillouin zone (BZ). This band structure is plotted in Fig.~\ref{fig:image3}(a), which also shows the path followed in the BZ.  Figure~\ref{fig:image3}(a) corresponds to the case when the  magnetization direction, $\textbf{\textit{M}}$, is  parallel to the $y$-direction, corresponding to the C$_{1h}$(C$_{1}$) magnetic group~\cite{dresselhaus2007group} [see Supplementary Note I].  The enhanced highlighted view of the topmost valence and lowest conduction bands [Fig.~\ref{fig:image3}(a)] shows valley polarization in both bands. The difference between the maxima along the $ K_{1}\to \Gamma$   and $\Gamma \to -K_{1}$ directions in the topmost valence band was found to be  $\sim40$\,meV. The lowest conduction band also shows a similar difference in the maxima, but with opposite sign ($\sim -38$\,meV).  This asymmetry in the band structure relative to the $\Gamma$-point comes from the interplay between the Rashba and exchange effects due to the broken space- and time-reversal symmetries ~\cite{PhysRevB.71.201403}.  In Fig.~\ref{fig:image3}(b), we plot spin-resolved band structure for 1H-YSSe, showing only the $\pm S_y$ spin projections, as the  $\pm S_x$ and $\pm S_z$  projections are almost negligible. Figure~\ref{fig:image3}(b) also shows an enhanced highlighted view of the topmost valence and lowest conduction bands, with colors encoding the spin projections. One can see that the topmost valence and lowest conduction bands are essentially spin up and spin down bands. This, in turn, implies that the spin, $\boldsymbol{S}$, is almost parallel to $\boldsymbol{M}$ in our case. This leads to a highly anisotropic Rashba spin-splitting, which is maximal when  $\boldsymbol{k}$  is perpendicular to spin $\boldsymbol{S}$ (or $\boldsymbol{M}$) and almost zero when $\boldsymbol{k}$ is parallel to $\boldsymbol{S}$ (or $\boldsymbol{M}$). Hence, the degree of the splitting in the maxima itself, depends on the component of the $\textbf{\textit{k}}$-vector perpendicular to the magnetization direction ($\textbf{\textit{M}}$ or $\textbf{\textit{S}}$) as can be seen from the Rashba Hamiltonian, $\mathcal{H}_{R}=\alpha_{R}(\boldsymbol{e}_{z} \times \boldsymbol{k}) \cdot \boldsymbol{S}$, which describes an electron of spin $\boldsymbol{S}$ moving with momentum $\boldsymbol{k}$ under the influence of an electric field oriented along the z-axis ($\boldsymbol{e}_{z}$)~\cite{PhysRevB.71.201403}. Here, $\alpha_{R}$ is the Rashba parameter. From the Rashba Hamiltonian, we find the splitting should be strongest along the $K_{1} \to \Gamma \to -K_{1}$  line as it is perpendicular to $\textbf{\textit{M}}$. This can be seen in the highlighted topmost valence and lowest conduction bands plotted in Figs~\ref{fig:image3}(a) and (b), which shows that the splitting in the maxima along the $ K_{1}\to \Gamma$ and $\Gamma \to -K_{1}$ is greater than the splitting in the maxima along the $ K_{2} \to \Gamma$ and $\Gamma \to -K_{2}$ directions.

In order to understand the relatively small value of MAE (about $-122\,\mu$eV) exhibited by our system, we considered the differences in the Bloch eigenstates: $\Delta E=\epsilon (\boldsymbol{k})$(010)- $\epsilon (\boldsymbol{k})$(001), calculated for two magnetic states with the magnetization $\boldsymbol{M}$ aligned in the $y$ and $z$-directions [see Fig.~\ref{fig:image3}(c)]. In this figure, only the energy differences involving the highest valence and lowest conduction bands are shown and the $\boldsymbol{k}$-lines are the same as those in Figs~\ref{fig:image3}(a) and (b).  Both the curves exhibit a counterbalanced trend between $-K_{1}-\Gamma$ and  $\Gamma- K_{1}$,  within the  $K_{1}-K_{2}$ region  and  between $ K_{2}-\Gamma$ and $\Gamma- (-K_{2})$. Moreover, the curves additionally exhibit a counterbalanced  trend with respect to each other: when the energy  difference for the valence band drops the corresponding difference for the conduction band rises and vice versa. Both these trends  are  explained by the (anisotropic) Rashba effect in the presence of ferromagnetic ordering. The counterbalanced contribution to the MAE in the $\boldsymbol{k}$-space, shown in Fig.~\ref{fig:image3}(c), corresponds to the hidden first order perturbation. Hence, our system exhibits what is known as ``unconventional MAE contributions in the  $\boldsymbol{k}$-space"~\cite{C7NR03164E}. When integrated over the entire BZ, the first order contribution to the MAE vanishes (being an odd function in the k-space), and the resulting MAE will be given by the relatively small second-order perturbation terms~\cite{MAE_2ndOrder_PRB.47.14932}, explaining the small value of MAE [see Supplementary Note II]. 


\begin{figure*}[ht]
	\centering
	\hspace{-0.5cm}
	\includegraphics[width=0.95\textwidth]{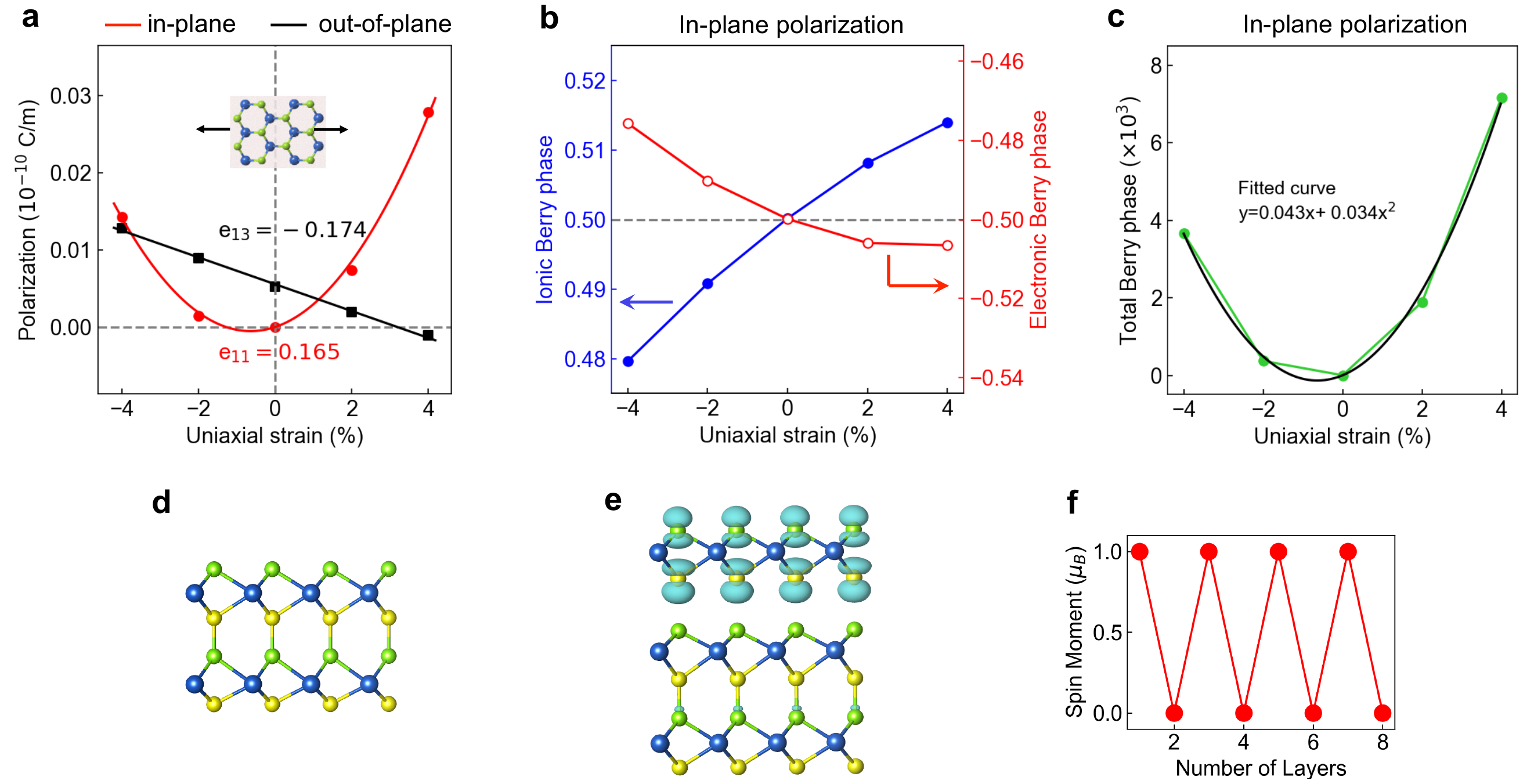}
	\caption{Strain and layer-thickness dependent properties of 1H-YSSe:  \textbf{a} Changes in in-plane and out-of-plane polarization of YSSe under the application of uniaxial strain along the armchair direction. \textbf{b} Electronic and ionic Berry phases (different scales) corresponding to the in-plane electric polarizations of the 1H-YSSe monolayer as a function of uniaxial strain along the armchair direction, showing exact cancellation of the two phases at 0\% strain and the switching of their role as the dominant contributor to the total phase as we go from lattice contraction to expansion. \textbf{c} Total Berry phase, showing the best fit curve (black line). \textbf{d} Lowest energy configuration (AA) of a bilayer of YSSe, wherein the S- and Se-atoms within the two layers bond, resulting in a non-magnetic structure. \textbf{e} Spin density plot ($\Delta_\rho=\rho^{Majority}-\rho^{Minority}$), showing that in the odd number of layers, the structure retains a net magnetic moment. \textbf{f} Magnetic moment of a stack of YSSe as a function of layer-thickness, showing odd-even layer-dependence.}
	\label{fig:image4}
\end{figure*}

\noindent \textbf{Piezoelectric properties of 1H-YSSe.} In addition to being a magnet, 1H-YSSe also displays piezoelectric properties owing to its noncentrosymmetric structure. The piezoelectric effect refers to the electromechanical coupling in which polarization changes in response to the applied strain. Due to the $C_{3v}$ symmetry of YSSe, the piezoelectric tensor has only two independent components: $e_{11}$  and $e_{13}$, associated with the in-plane and out-of-plane polarizations, respectively.
We calculated the polarization of YSSe using the Berry phase method~\cite{King-Smith1993,Resta2007} (see Methods section). We find that in the unstrained state, there is no in-plane polarization (as required by $C_{3v}$ symmetry), while there is a spontaneous polarization of $0.524\times 10^{-12}$\,C/m along the symmetry-allowed $C_{3}$-axis ($z$-direction), with the dipole moment directed from the Se to S atoms. Figure~\ref{fig:image4}(a) shows the in-plane and out-of-plane electric polarizations, as a function of the uniaxial strain along the armchair direction. Here we define the strain percentage in a system with lattice constant $b$ along the armchair direction as: $100 \times (b-b_{0})/b_{0}$, with $b_{0}$ being the equilibrium lattice constant. The slopes of the curves in Fig.~\ref{fig:image4}(a) give $e_{11}=0.165\times10^{-10}$\,C/m (in-plane piezoelectric coefficient at 0\% strain), and $e_{13}=-0.174\times10^{-10}$\,C/m (out-of-plane piezoelectric coefficient), respectively.  The best-fit curve for the highly nonlinear behaviour of the in-plane  polarization is given by $b_{1}x+b_{2}x^2$, where the linear coefficient $b_1 \equiv e_{11}$. The quadratic coefficient, $b_{2}=0.130\times10^{-10}$\,C/m, is unusually large as compared to those calculated for other similar 2D materials, such as MoSSe [see Supplementary Fig. 2].

In order to understand the origin of the non-linear behaviour of the in-plane polarization as a function of strain, we analyzed both the electronic and ionic contributions to the total Berry phase, $\phi_{tot}$.  Figure~\ref{fig:image4}(b) shows the electronic  and the ionic phases, which are of opposite signs, resulting in very small values for $\phi_{tot}$.  In fact, at 0\% strain the values of the electronic and ionic phases are exactly half of the polarization quantum each ($-0.5$ and $0.5$, respectively), and they cancel each other exactly, giving $\phi_{tot}=0$. In going from $0\%$ strain to  $+4\%$ (corresponding to the lattice expansion along the armchair direction), the electronic phase changes by a mere $0.665\times10^{-2}$, while the ionic phase changes by $-1.382\times10^{-2}$. Hence, the ionic contribution dominates the total phase in this region [see Fig.~\ref{fig:image4}(b)]. In particular, in the strain region between $+2\%$ and $+4\%$, the electronic phase is almost a constant. This means that the charge centers of the maximally-localized Wannier functions practically follow a homogeneous deformation~\cite{Vanderbilt2000}. At the same time, the ionic positions fail to follow a homogeneous deformation (i.e. they undergo internal distortions relative to the lattice). Hence, we observe an overwhelming dominance of the ionic contribution to the total phase. The situation changes in going from $0\%$ to $-4\%$, which corresponds to the lattice contraction along the armchair direction. Here, in contrast to the case of lattice expansion, it is the change in the electronic phase ($=-2.418\times10^{-2}$) that dominates in the total phase (as compared to the ionic phase that changes by $2.052\times10^{-2}$). Thus, 0\% strain serves as a crossover point, across which the role of the dominant contributor to total phase is switched between the ionic and electronic phases. This effect, which is especially enhanced by the near-cancellation of the electronic and ionic terms in the total phase, is the reason for the nonlinear piezoelectricity in the system.

\noindent \textbf{Layer-thickness dependent properties of 1H-YSSe.} Lastly, we considered the influence of layer-thickness on the electronic structure properties of a YSSe monolayer.  The monolayers of YSSe can be stacked in different registries such as: (i) AA-stacking, with the chalcogen of one layer over the chalcogen of the next layer, and (ii) AB stacking, with the chalcogen in one layer over the metal atom of the second layer [see Supplementary Fig. 3]. Unlike most TMDs, YSSe prefers AA stacking over AB, as this stacking allows YSSe to form bonds between the two layers, lowering its energy by about 1.35\,eV  [see Fig.~\ref{fig:image4}(d)].  The resultant bilayer structure is non-magnetic.  On the other hand, the trilayer structure is again magnetic, with a moment of $1\mu_{B}$.  In a trilayer, two of the layers form bonds, allowing the third non-bonded layer to retain its magnetism. This can be seen in Fig.~\ref{fig:image4}(e), which shows the spin-density plot ($\Delta_\rho=\rho^{Majority}-\rho^{Minority}$) for a trilayer. We find that YSSe shows this interesting odd-even layer dependence of magnetic properties for all tested thicknesses [see Fig.~\ref{fig:image4}(f)].

\vspace{12pt}
\noindent \textbf{Discussion}
\vspace{8pt}

Using first principles based-methods, we have predicted a new magnetic Janus TMD, 1H-YSSe. Due to its unique structural, electronic and spin properties, 1H-YSSe displays an interesting and uncommon combination of magnetism and piezoelectricity in a single material.  Such materials are rare because conventional magnets have localized electrons in partially filled $3d$-orbitals of transition metals (or the $4f$-orbitals of rare earth metals), while piezoelectric materials are insulators, usually consisting of $d^{0}$ or $d^{10}$ elements~\cite{Spaldin2000}. A material where both of these properties coexist is potentially useful in proximity effect devices, such as the superconductor/semiconductor/magnet heterostructures being explored as solid-state platforms for Majorana bound states, where the latter can be tuned/controlled by external stimuli (e.g. electric field and/or strain) applied to the magnet.  Furthermore, the broken space- and time-symmetry ensures valley polarization in the topmost valence and the lowest conduction bands. These properties make YSSe a promising material for novel applications such as ultra-compact spintronics, valleytronics, and quantum devices.
\vspace{12pt}

\textbf{Methods}

We carried out the spin-polarized  density functional theory (DFT) calculations using the projected augmented wave method (PAW)~\cite{Kresse1996a, Kresse1996b} as implemented in the Vienna \textit{Ab-Initio} Simulation package.  A Monkhorst-Pack~\cite{kpoint2} k-point mesh of size 16$\times$16$\times$1 and a plane wave energy cutoff of 550\,eV were used in the calculations.  We employed the generalized gradient approximation of Perdew, Burke and Ernzerhof (PBE)~\cite{Perdew1996} for the exchange-correlation functional.  In the Supplement (Supplementary Fig. 4), we also provide the results of the computationally-expensive hybrid functional, HSE06~\cite{HSE03, HSE06}, which includes a fixed percentage of Hartree-Fock exchange in the exchange-correlation functional. This functional was used solely for the semiconducting phase of YSSe for its greater accuracy in predicting band gaps of semiconductors. For the case of the metallic phase of YSSe, we used only PBE, as hybrid functionals are known to give erroneous results for metals~\cite{Peihong_HSE_metal_2016}.  The structure was optimized by ensuring that the forces on each atom are less than 0.001\,eV/\AA{}.  We also ensured that the interaction between a monolayer and its images is minimized by adding a vacuum layer of 20\,\AA{}. The dynamical and thermodynamical stability of the YSSe monolayer was ensured by calculating phonon dispersion using the finite displacement method~\cite{Parlinski1997} as implemented in the Phonopy code~\cite{togo2008} and by performing \textit{ab-initio} molecular dynamic simulations using canonical ensembles. 

The piezoelectric properties of YSSe were obtained via the Berry-phase approach~\cite{King-Smith1993,Resta2007}. To calculate the Berry phases, we used the Ceperley-Alder functional~\cite{Ceperley1980} and Hartwigsen-Goedecker-Hutter pseudopotentials~\cite{Hartwigsen1998} as implemented in ABINIT package~\cite{Gonze2009}. The  chosen pseudopotentials are characterized by 6, 3 and 6 valence electrons, for S, Y, and Se ions, respectively. 
Strictly speaking, the existence of magnetic ordering in the system can additionally induce a spontaneous in-plane polarization via spin-orbit coupling, as in the  case of terbium manganite~\cite{Malashevich2009}. This magneto-electric coupling, however, was found to be negligibly small. Hence, we did not take into account any magnetic effects on electric polarizations and assumed that the system has a point group of $C_{3v}$ (i.e. the symmetry is not reduced by magnetism).


%

\vspace{12pt}


\noindent \textbf{\large{Acknowledgments}}

\noindent This work is supported by the W. M. Keck Foundation Research Award and the National Science Foundation (under NSF grant number DMR-1752840 and the STC Center for Integrated Quantum Materials under NSF Grant No. DMR-1231319). The computational support is provided by the Extreme Science and Engineering Discovery Environment (XSEDE) under Project PHY180014, which is supported by National Science Foundation grant number ACI-1548562. For three-dimensional visualization of crystals and volumetric data, use of VESTA 3 software is acknowledged.

\vspace{12pt}

\noindent \textbf{\large{Author Contributions}}
\vspace{8pt}

\noindent  P.D. conceived and designed the project. P.K., I.N. and P.D. equally contributed to the first principles calculations performed to determine the properties of YSSe monolayers. P.M. determined properties of multiple layers of YSSe.  All authors were involved in the analysis of results and discussions. P.K. I.N. and P.D. wrote the manuscript. All authors reviewed and contributed to the final revision of the manuscript.

\vspace{8pt}

\noindent \textbf{\large{Additional information}}
\vspace{12pt}

\noindent \textbf{Competing Interests} 
\vspace{4pt}

\noindent The author declares no competing financial interests.

\vspace{4pt}

\noindent \textbf{Correspondence} 
\vspace{4pt}

\noindent Correspondence and requests for materials should be addressed to P.D.~(email: pratibha.dev@howard.edu).
\vspace{12pt}

\end{document}